\documentstyle[12pt,aasms4]{article}

\begin{document}

\title{Where is SGR1806-20?}

\author{K. Hurley}
\affil{University of California, Berkeley, Space Sciences Laboratory,
Berkeley, CA 94720-7450}
\authoremail{khurley@sunspot.ssl.berkeley.edu}
\author{C. Kouveliotou} 
\affil{Universities Space Research Association at NASA Marshall Space Flight Center, 
ES-84, Huntsville AL 35812}
\author{T. Cline}
\affil{NASA Goddard Space Flight Center, Greenbelt, MD 20771}
\author{E. Mazets, S. Golenetskii, D. D. Frederiks}
\affil{Ioffe Physical-Technical Institute, St. Petersburg, 194021, Russia}
\author{J. van Paradijs\altaffilmark{1}, University of Alabama in Huntsville, AL 35899}
\altaffiltext{1}{Astronomical Institute `Anton Pannekoek',
University of Amsterdam, The Netherlands}

\begin{abstract}

We apply a statistical method to derive very precise locations for soft gamma
repeaters using data from the interplanetary network.  We demonstrate the validity
of the method by deriving a 600 arcsec$^{2}$ error ellipse for SGR1900+14 whose
center agrees well with the VLA source position.  We then apply it to SGR1806-20, for
which we obtain a 230 arcsec$^{2}$ error ellipse, the smallest burst error
box to date.  We find that the most likely
position of the source has a small but significant displacement from that of
the non-thermal core of the radio supernova remnant G10.0-0.3, which was
previously thought to be the position of the repeater.  We propose a different
model to explain the changing supernova remnant morphology and the positions of the luminous blue variable
and the bursting source.  

\end{abstract}


\section{Introduction}

The four known soft gamma repeaters (SGRs) are neutron stars in or near radio or optical supernova remnants.  SGR1806-20 was discovered in 1986 (Laros et al. 1986) and underwent a period of intense activity in 1987 (Laros et al. 1987, Kouveliotou et al. 1987) which led to its localization
to an $\sim$ 400
arcmin$^{2}$ error ellipse (Atteia et al. 1987).  Based on this position, Kulkarni \& Frail (1993) suggested that the SGR was associated with the Galactic radio supernova remnant (SNR) G10.0-0.3.  This was confirmed when the ASCA spacecraft observed and imaged the source in outburst, leading to a $\sim$ 1' radius error circle (Murakami et al. 1994).  ROSAT observations of the quiescent X-ray source associated with SGR1806-20 confirmed the ASCA data (Cooke 1993; Cooke et al. 1993).  It is believed that the SGRs are 'magnetars', i.e. single neutron stars in which the magnetic field energy dominates all other sources of energy, including rotation (Duncan \& Thompson 1992).  In the case of SGR1806-20, evidence for this model comes from observations of the period and period derivative of the quiescent soft X-ray emission (Kouveliotou et al. 1998). 
 
Studies of the radio nebula show evidence for changes in its morphology on $\sim$ year timescales, and suggest that the neutron star may be located at the non-thermal core of the radio emission (Frail et al. 1997).  The position of the core also coincides with that of an unusual star, identified as a luminous blue variable (LBV) by van Kerkwijk et al. (1995).  This
appears to be the only case so far of an SGR with an optical stellar counterpart, and the connection between this object and the SGR has been unclear up to now.

SGR1806-20 has remained active over the past several years, and many bursts have been detected by the Interplanetary Network (IPN), consisting primarily in this case of BATSE,  \it Ulysses \rm, and KONUS-WIND.  However, only eight events have been intense enough to trigger both  \it Ulysses \rm and a near-Earth spacecraft, resulting in high time resolution data (the other bursts were recorded with lower time resolution by one or more instruments).    It is these triggered events which lead to the most precise determination of the source position by triangulation.

\section{Observations}

Details of the eight triggered bursts are given in Table 1.  In each case, triangulation using \it Ulysses \rm and either BATSE or KONUS results in a single annulus of width $\sim$23 - 28" which defines the possible arrival direction for the burst.  Two such annuli define an error box, if the angular separation between their centers is sufficient to prevent the annuli from intersecting at grazing incidence.  Over the $\sim$2 yr period analyzed here, the  \it Ulysses \rm-Earth vector moved sufficiently to define a non-degenerate error box.  With three or more annuli, the problem of defining the source location becomes overdetermined, and we can use a statistical method to derive the most probable source location.  This consists of defining a chisquare which is a function of an assumed source position in right ascension and declination, and of the parameters describing the eight annuli.  Let 
$\alpha, \delta$ be the right ascension and declination of
the assumed source position, and let $\alpha_i, \delta_i, \theta_i$
be the right ascension, declination, and radius of the ith annulus.
Then the angular distance d$_i$ between the two is given by
\begin{equation}
d_i= \theta_i - \cos^{-1}(\sin(\delta) \sin(\delta_i) +
\cos(\delta) \cos(\delta_i) \cos(\alpha - \alpha_i) )  
\end{equation}.
If the 1 $\sigma$ uncertainty in the annulus width is $\sigma_i$, then
\begin{equation}
\chi^{2}=\sum_{i}\frac{d_i^2}{\sigma_i^2}.
\end{equation}
The assumed source position is varied to obtain a minimum chisquare; 1, 2, and 3 $\sigma$
equivalent confidence contours in $\alpha$ and $\delta$ are found by increasing
$\chi^{2}_{min}$ by 2.3, 6.2, and 11.8.

We have tested this method on six IPN annuli for SGR1900+14 (Table 2),
whose precise (sub-arcsecond) location is
known from VLA observations of a particle outburst (Frail, Kulkarni, and Bloom 1999)
following the giant flare of 1998 August 27 (Hurley et al. 1999a).  The result is
shown in figure 1.  The 3$\sigma$ error ellipse has an area of $\sim$600 arcsec$^{2}$,
and the best fitting position for the SGR, at $\rm \alpha(2000)=19^h 07^m 14.3^s, \delta(2000)=
9^o 19\arcmin 19\arcsec$, has a $\chi^2$ of 1.05 for 4 degrees of freedom
(six annuli, minus the two fitting parameters $\alpha, \delta$).  It lies $\sim0.6 \arcsec$
from the VLA position.

The results of applying the method to SGR1806-20 are shown in figure 2.
The best fit position is at $\rm \alpha(2000)=18^h 08^m 39.4^s, \delta(2000)=
-20^o 24\arcmin 38.6\arcsec$, with a $\chi^2$ of 3.35 for 6 degrees of freedom
(8 annuli minus two fitting parameters).  It lies $\sim$15 \arcsec from the center
of the non-thermal core, and well outside it.  The 3$\sigma$ error ellipse has an
area of $\sim$ 230 arcsec$^{2}$, making it the smallest burst error box determined
to date (the 324 arcsec$^{2}$ error box of the
1979 March 5 burst was, until now, ``the most precisely determined gamma-ray source error box
in existence'' - Cline et al. 1982).  The position of the non-thermal core has a 
total $\chi^2$ of 101.  

\section{Accuracy of the Method}

Since each individual annulus gives, in effect, an underdetermined source
position, it is possible in principle that unknown systematic errors might
affect the location accuracy.  For example, timing errors of 96 to 206 ms
in the \it Ulysses \rm data could shift the positions of the annuli by different
amounts and
make them all consistent with that of the non-thermal core.  Apart from the unlikely
combination of errors which this would require (i.e., each annulus would
have to be subject to a different error in such a way as to make the
erroneous best fit position have an acceptable $\chi^2$), there are several independent
confirmations of the accuracy of the triangulation method.  The first is the
excellent agreement between the VLA and triangulated positions of SGR1900+14.
The second is the agreement between IPN positions and the positions of gamma-ray bursts
with optical counterparts (e.g. Hurley et al. 1997).  The third and most stringent,
however, is the confirmation of the \it Ulysses \rm spacecraft timing and ephemeris
by end-to-end timing tests, in which commands are sent to the GRB experiment
at precisely known times, and the times of their execution onboard the spacecraft 
are recorded and compared with the expected times.  Because of command buffering
on the spacecraft, there are random delays in the execution of these commands,
and the timing is verified to different accuracies during different tests.
However, the tests before, during, and after the eight bursts in Table 1 took
place on 1996 October 1, 1997 February 19, 1997 August 25, 1998 February 18, 1998
August 21, and 1999 March 7, and indicated that the timing errors at those times
could not exceed 19, 21, 39, 29, 112, and 1 ms respectively.  For comparison,
the 3 $\sigma$ uncertainties in these triangulations have been taken to be
125 ms.  This includes both the statistical errors, and a conservative estimate of 
unknown timing and spacecraft ephemeris errors.  The low $\chi^2$ values for
the two SGR positions are probably due in part to this estimate.  Thus the most
likely explanation of our results is indeed that SGR1806-20 is not in the non-thermal
core of G10.0-0.3, as has been assumed up to now. In this respect, SGR1806-20
resembles SGR1627-41, which also displays a significant displacement from the
core of its radio SNR (Hurley et al. 1999b).
 
\section{Discussion}

The association of van Kerkwijk et al.'s (1995) possible LBV with the SNR is compelling;
they estimate that there are only several hundred stars this luminous in the galaxy,
and this one lies within 1 $\arcsec$ of the radio peak.  Indeed, in assuming a
distance to the object of 6 kpc, they may have
underestimated the star's luminosity; a better distance estimate is now 14.5 kpc (Corbel
et al. 1997), giving a luminosity of $\rm 6 \times 10^6 L_{\odot}$.  The fact that this
object has not yet been observed to vary is not an argument against the LBV identification:
Humphreys \& Davidson (1994) note that LBV's do not
\it always \rm appear blue or variable.  They are simply very luminous, unstable hot
supergiants which undergo irregular eruptions.  In a giant eruption, they may radiate
as much luminous energy as a supernova, and eject a solar mass of material.  But this
does not explain the SGR bursts, the changing radio morphology,  or the displacement between the radio core and the source of the bursts.

We propose that the LBV drives the morphological changes.   LBV's are characterized by sporadic mass loss
rates of up to $\rm \sim 10^{-4}M_{\odot}/y$ (Humphreys \& Davidson 1994) and more.  Moreover,
these flows may be bipolar or jetlike, as in the case of $\eta$ Car or P Cygni
(Meaburn, Lopez \& O'Connor 1999).  Van Kerkwijk et al's (1995) measurements
of the possible LBV in G10.0-0.3 indicate an outflow velocity of 500 km/s.  Coupled with
a mass loss rate of $\rm 10^{-4}M_{\odot}/y$, this gives a total wind energy of $\rm 2.5 \times
10^{44} erg/y$, or a factor of $\sim$30 greater than the rate of energy deposition 
into the radio nebula by the neutron
star in the model of Frail et al (1997).  Thus the LBV is easily capable of supplying the
energy to explain the changing radio morphology.  In the case of $\eta$ Car, the LBV
not only changes the morphology of its radio nebula dramatically, but it also powers
the (apparently non-thermal) radio nebula (Duncan et al. 1995; Duncan, White, \& Lim 1997).

We believe that the magnetar model is the best current explanation for the bursts.
It is possible that the SGR is not associated with the radio nebula, and that we
are simply observing a chance alignment of the two.
But if the two are indeed associated, the SGR and the LBV may
once have formed a binary system, which became unbound following the supernova
explosion.  In the magnetar model, the neutron star may be born with a kick velocity $>$1000 km/s (Duncan \& Thomson 1992).    If we assume that the distance to SGR1806-20 is 14.5 kpc (Corbel et al. 1997), that its age is 10,000 y, and that
the neutron star originated at the position of the LBV/non-thermal core, its approximate transverse velocity is a rather modest 100 km/s.  (This estimate is subject to large uncertainties
due to the unknown age of the SNR; also, the actual space velocity could be
much larger).  This certainly does not strain the magnetar model, but it does raise
another interesting question.  Why did the SGR progenitor form a neutron star rather
than a black hole, given that it must have been very massive to
end its life earlier than the LBV?  In any case, SGR1806 now appears to be similar
to the other SGRs in that there is no associated radio emission at its position,
except for the brief radio flare from SGR1900+14 (Frail et al. 1999).

\acknowledgments
KH is grateful to JPL for \it Ulysses \rm support under Contract 958056,
and to NASA for Compton Gamma-Ray Observatory support under
grant NAG 5-3811.  We thank the referee, M. van Kerkwijk, for helpful
comments, and P. Li for his analysis of the ROSAT data.

\newpage

\figcaption{Six IPN annuli (black lines), and the 1, 2, and 3 $\sigma$ equivalent 
confidence contours (red annuli) for SGR1900+14.  The best fit position and the
position of the radio source are also indicated. \label{fig1}}

\figcaption{Eight IPN annuli (black lines), and the 1, 2, and 3 $\sigma$ equivalent
confidence contours (red annuli) for SGR1806-20.  The best fit position and the position of
the non-thermal core are indicated.  The ASCA error circle is just visible in
the lower left and upper left hand corners; its radius is 1', quoted as a systematic error, with no confidence limit given (Murakami et al. 1994).  The ROSAT PSPC error circle is at the
center; its radius is 11", with no confidence limit quoted (Cooke et al. 1993).
We have reanalyzed the ROSAT data and confirm the presence of a weak source at
this position, but are unable to establish confidence limits for its position.  The 3.6 cm radio contours of G10.0-0.3 are also shown,
from Vasisht, Frail, \& Kulkarni (1995).   \label{fig2}
}

\clearpage
\begin{deluxetable}{cccccc}
\tablecaption{\it IPN Triggered Observations of SGR1806-20.  }
\tablehead{
\colhead{} & \colhead{} & \multicolumn{2}{c}{Annulus center} & \colhead{} & \colhead{} \\
\colhead{Date} & \colhead{UT, s}  & \colhead{$\rm \alpha(2000)$}& \colhead{$\rm \delta(2000)$} & \colhead{Radius, $\theta$} & \colhead{3$\sigma$ half-width} \\
\colhead{} & \colhead{} & \colhead{(deg.)} & \colhead{(deg.)} & \colhead{(deg.)}
 & \colhead{(deg.)}
}

\startdata
961119 	&	19826 & 358.8945 & -30.9318 & 76.9907 & 0.0031	\nl
961230	&	65965	& 359.0677 & -32.4563 &	76.7108 & 0.0035	\nl
970124	&	28043	& 355.4360 & -34.4572 &	73.2710 & 0.0038	\nl
970414	&	39219	& 334.4596 & -34.6283 &	56.1701 & 0.0039  \nl
970827	&	06045	& 341.3113 & -19.3428 &	64.5125 & 0.0026  \nl
970902	&	39347	& 342.2306 & -18.6910 &	65.5168 & 0.0026	\nl
980805	&	37673	& 335.3991 & -06.2919 &	62.7583 & 0.0026	\nl
990205	&	31452	& 341.2378 & +09.8529 &	74.3287 & 0.0035	\nl

\enddata
\end{deluxetable}

\clearpage
\begin{deluxetable}{cccccc}
\tablecaption{\it IPN Triggered Observations of SGR1900+14.  }
\tablehead{
\colhead{} & \colhead{} & \multicolumn{2}{c}{Annulus center} & \colhead{} & \colhead{} \\
\colhead{Date} & \colhead{UT, s}  & \colhead{$\rm \alpha(2000)$}& \colhead{$\rm \delta(2000)$} & \colhead{Radius, $\theta$} & \colhead{3$\sigma$ half-width} \\
\colhead{} & \colhead{} & \colhead{(deg.)} & \colhead{(deg.)} & \colhead{(deg.)}
 & \colhead{(deg.)}
}

\startdata
980526 	&	80649 & 329.5855 & -11.4558 & 47.3345 & 0.0037	\nl
980530	&	84457	& 329.6579 & -11.2371 &	47.3091 & 0.0036	\nl
980719	&	30383 & 333.3281 &  -7.7778 &	49.3927 & 0.0031	\nl
980901	&	58382	& 339.0121 &  -3.7061 &	53.6270 & 0.0029  \nl
981022	&	56447	& 345.2596 &  +1.5599 &	58.6313 & 0.0029	\nl
981028	&	83020	& 345.8533 &  +2.2211 &	59.1029 & 0.0029	\nl

\enddata
\end{deluxetable}

\end{document}